\documentclass[pra,aps,showpacs,twocolumn]{revtex4}
\newcommand{\ket}[1]{|#1\rangle}
\newcommand{\bra}[1]{\langle #1|}
\newcommand{\Tr}{\text{Tr}}
\begin{document}
\title{Mixed state non-Abelian holonomy for subsystems}
\author{Mikael Nordling and
Erik Sj\"{o}qvist\footnote{Electronic address: erik.sjoqvist@kvac.uu.se}}
\affiliation{Department of Quantum Chemistry, Uppsala University,
Box 518, Se-751 20 Uppsala, Sweden.}
\date{\today}
\begin{abstract}
Non-Abelian holonomy in dynamical systems may arise in adiabatic
transport of energetically degenerate sets of states. We examine
such a holonomy structure for mixtures of energetically degenerate
quantal states. We demonstrate that this structure has a natural
interpretation in terms of the standard Wilczek-Zee holonomy
associated with a certain class of Hamiltonians that couple the system
to an ancilla. The mixed state holonomy is analyzed for holonomic
quantum computation using ion traps.
\end{abstract}
\maketitle
\section{Introduction}
Non-Abelian holonomy in simple dynamical systems \cite{wilczek84} has
found application in a variety of contexts, such as nuclear rotations
of diatoms \cite{moody86,jackiw86}, molecular Kramers doublets
\cite{mead87,mead92}, nuclear quadrupole resonance (NQR)
\cite{tycko87,zee88,zwanziger90}, dynamics of deformable bodies
\cite{shapere89a,shapere89b}, the n-body problem \cite{littlejohn97},
semiconductor heterostructures \cite{arovas98}, and ion traps
\cite{unanyan99}. Recently, it has been pointed out
\cite{zanardi99,pachos00} that non-Abelian holonomy may be
used in the construction of universal sets of quantum gates for the
purpose to achieve fault tolerant quantum computation. This has
triggered further work on holonomy effects for quantum computation
\cite{duan01,pachos02,faoro03,choi03,cholascinski04,zhu04,cen03,
niskanen03,tanimura03} and quantum information 
\cite{zanardi01,fuentes02,marzlin03,li04}.

An important issue for holonomic quantum information processing is its
robustness to imperfections, such as decoherence and random unitary
perturbations. The effect of imperfections on the quantum gates has
been analyzed \cite{pachos02,fuentes03,solinas03,cen04}, supporting
the alleged robustness of holonomic quantum computation. It remains
however to address the appearance of non-Abelian holonomy related to
the motion of the quantal states themselves. Here, we wish to
address this issue by putting forward an attempt to define a concept
of non-Abelian holonomy in terms of incoherent mixtures of
energetically degenerate quantal states.

The main idea of this paper is to relate the mixed state time
evolution of a system $S$ to the holonomic transformation of a certain
class of pure state systems consisting of $S$ coupled to an ancilla
system $A$. For the whole system $S+A$, this coupling creates in a
purely geometric fashion an entangled state whose partial trace over
$A$ corresponds in general to a nonpure state of $S$. From this
perspective, we may regard the present analysis as the identification
of a class of $S-A$ interaction Hamiltonians having degenerate
eigenspaces that admit a non-Abelian holonomy structure for the mixed
states of $S$. This structure reduces to the Wilczek-Zee pure state
holonomy \cite{wilczek84} of the $S$ system when the $S-A$ interaction
vanishes. We examine the present type of mixed state holonomy in the
case where the pure state limit corresponds to the ion trap system in
Refs. \cite{unanyan99,duan01}.

\section{Mixed state holonomy}
\label{sec:holonomy}
Let a quantal system $S$ be exposed to the Hamiltonian $H(q)$, $q$
being some external control parameters that vary around a closed
path ${\cal C}: t\in [0,T] \rightarrow q_t$ in parameter space.
Let ${\cal H}_S$ be the Hilbert space of $S$ and let ${\cal M} (q)$
denote the linear span of the $N$-fold degenerate instantaneous
eigenstates $\ket{\xi_a (q)},a=0, \ldots, N-1$, $N \leq \dim
{\cal H}_S$, of $H(q)$. Under the condition that $T$ is large
enough so that the adiabatic theorem holds, any initial state
$\ket{\psi (0)} \in {\cal M} (q_0)$ is mapped to $\ket{\psi (t)}
\in {\cal M} (q_t)$. Thus,
\begin{eqnarray}
\ket{\psi (0)} & = & \ket{\xi_a (q_0)} \rightarrow 
\nonumber \\ 
\ket{\psi (t)} 
 & = & \exp \left( -i \int_0^T E(q_t) dt \right) 
\nonumber \\ 
 & & \sum_b U_{ab}(t) \ket{\xi_b (q_t)}
\end{eqnarray}
with $U(t)$ unitary, $E(q_t)$ the common instantaneous energy of
the degenerate subspace, and we have put $\hbar =1$ from now on.
It follows that
\begin{eqnarray}
U[{\cal C}] =
{\bf P} \exp \left( - \oint_{{\cal C}} A \right)
\end{eqnarray}
is the holonomy transformation for ${\cal C}$.  Here,
${\bf P}$ is path ordering and $A = \sum_{\mu} A_{\mu}(q)
dq^{\mu}$ is the connection one-form with
\begin{eqnarray}
A_{ab,\mu}(q) = \bra{\xi_b(q)} \partial_{\mu} \ket{\xi_a(q)}
\label{eq:pureWZ}
\end{eqnarray}
the components of the Wilczek-Zee gauge potential and
$\partial_{\mu} = \partial /\partial q_{\mu} $.

We now apply the above Wilczek-Zee prescription in order to achieve
the main idea of this paper, which is to propose a concept of mixed state 
non-Abelian holonomy for subsystems. To this end, we first identify a
proper set of degenerate entangled energy eigenstates by considering
the extended Hilbert space $\tilde{{\cal H}} = {\cal H}_{S} \otimes
{\cal H}_{A}$, where ${\cal H}_A$ is an ancilla Hilbert space taken to
be identical to ${\cal H}_S$ and the vectors in $\tilde{{\cal H}}$
being purifications of the mixed states of $S$. The essential
criterion is that this set should define the standard pure state
holonomy of the $S$ system in the product state limit. This may be
achieved by letting $\tilde{H} (q)$ be the $N$-fold degenerate
Hamiltonian of the composite system $S+A$ and let the members of the
instantaneous set of orthonormal eigenvectors $\{ \ket{\Xi_a (q)} \}$,
spanning the degenerate subspace of $\tilde{H}(q)$, be iso-entangled,
i.e., that they correspond to purifications of mixed states $\Tr_A
\ket{\Xi_a (q)} \bra{\Xi_a (q)}$ all having the same eigenvalues
$\lambda_k$, $k=0,\ldots,N-1$. If $\lambda_l \rightarrow 1$ and
$\lambda_{k\neq l} \rightarrow 0$, then it follows that $\{ \ket{\Xi_a
(q)} \} \rightarrow
\{ \ket{\xi_a (q)} \otimes \ket{\varphi} \}$, for some 
$\ket{\varphi} \in {\cal H}_A$, which establishes the 
desired connection to the holonomy in the pure state case. 

Now, the iso-entangled energy eigenstates may explicitly be 
taken as
\begin{equation}
\ket{\Xi_a (q)} =
\sum_{k=0}^{N-1} \sqrt{\lambda_k} \ket{\xi_{[a+k]_N}(q)} 
\otimes \ket{\varphi_k},
\end{equation}
where $[a+k]_N \equiv a+k \ \textrm{mod} \ N$. For this orthonormal 
set, we obtain the gauge potential as
\begin{eqnarray}
A_{ab,\mu} & = & \bra{\Xi_b (q)} \partial_{\mu} \ket{\Xi_a (q)} 
\nonumber \\ 
 & = & \sum_{k=0}^{N-1} \lambda_k \bra{\xi_{[b+k]_N}(q)} 
\partial_{\mu}\ket{\xi_{[a+k]_N}(q)}.
\label{eq:msconnection}
\end{eqnarray}
The resulting unitarity $U[{\cal C}]$ for the loop ${\cal C}$ 
in parameter space is the desired mixed state non-Abelian holonomy 
of the set of energetically degenerate reduced density operators 
$\Tr_A \ket{\Xi_a (q)} \bra{\Xi_a (q)}$ pertaining to the subsystem $S$. 

To further analyze the physical meaning of the proposed form of mixed
state holonomy, let us note that the Hamiltonian corresponding to this
setup may be written as
\begin{eqnarray}
\tilde{H} (q) & = & E (q) \sum_a \ket{\Xi_a (q)} \bra{\Xi_a (q)} +
\Delta H(q) 
\nonumber \\ 
 & \equiv & H_E (q) +\Delta H(q) ,
\end{eqnarray}
where $E (q)$ is the energy of the degenerate subspace and $\Delta
H(q)$ acts only on the orthogonal complement. Thus, $\tilde{H} (q)$ 
induces interaction between $S$ and $A$ along the path ${\cal C}$. 
For example, in the $N=2$ case, we have
\begin{eqnarray}
H_E & = & E(q) \big( \sigma_0^S(q) \otimes \left[ \lambda_0 \ket{\varphi_0}
\bra{\varphi_0} + \lambda_1 \ket{\varphi_1} \bra{\varphi_1} \right] 
\nonumber \\ 
 & & + \sqrt{\lambda_0\lambda_1} \sigma_1^S(q) \otimes \sigma_1^A\big) ,
\label{eq:NH2D}
\end{eqnarray}
where
\begin{eqnarray}
\sigma_0^S (q) & = & \ket{\xi_0(q)} \bra{\xi_0(q)} +
\ket{\xi_1(q)} \bra{\xi_1(q)} ,
\nonumber \\
\sigma_1^S (q) & = & \ket{\xi_0(q)} \bra{\xi_1(q)} +
\ket{\xi_1(q)} \bra{\xi_0(q)} ,
\nonumber \\
\sigma_1^A & = &
\ket{\varphi_0} \bra{\varphi_1} + \ket{\varphi_1} \bra{\varphi_0} .
\end{eqnarray}
In the pure state limit $\lambda_0 \lambda_1 \rightarrow 0$ we 
have $\ket{\Xi_a(q)} = \ket{\xi_a(q)} \otimes \ket{\varphi_0}$,
$\forall a$, which decouples the Hamiltonian and leads to the
pure state Wilzcek-Zee gauge potential in Eq. (\ref{eq:pureWZ}).

Now, since $S$ in general interacts with $A$, one should expect 
that the evolution of the density operator $\rho$ is nonunitary. 
To check this, one may assume that $\rho (0) = \Tr_A \ket{\psi (0)} 
\bra{\psi (0)}$, with $\ket{\psi (0)} = \ket{\Xi_a (q_0)} \in 
{\cal H}_S \otimes {\cal H}_A$, calculate the final density 
operator as
\begin{eqnarray}
\rho (T) & = & \Tr_A \ket{\psi (T)}\bra{\psi (T)} 
\nonumber \\ 
 & = & \sum_{b,c} U_{ab}^{\ast} [{\cal C}] U_{ac} [{\cal C}]
\Tr_A \ket{\Xi_c(0)}\bra{\Xi_b(0)} ,
\end{eqnarray}
and compare the initial and final degree of mixing for instance
by computing $\Tr \rho^2$. Due to the appearance of off-diagonal
terms in $\rho (T)$, it is indeed apparent that $\Tr \rho^2(T)
\neq \Tr \rho^2 (0)$, in general. 

\section{Physical example}
\label{sec:examples}

A scheme to realize geometric quantum computation for ion traps 
has been proposed \cite{duan01} (see also Ref. \cite{unanyan99}). 
Here we extend the $U_2$ gate in Ref. \cite{duan01}, using our 
gauge potential for mixed states.

Suppose we have an ion with three metastable ground states denoted by
$\ket{0}$, $\ket{1}$, and $\ket{a}$, and one excited state $\ket{e}$. A
laser couples the ground states with the excited state, as described by
the Hamiltonian
\begin{eqnarray}
\label{eq:ITHam}
H = \ket{e} \big( \bra{0} \omega_0 +
\bra{1} \omega_1 + \bra{a} \omega_a \big) + \textrm{h.c.} ,
\end{eqnarray}
where $\omega_0,\omega_1,\omega_a$ are complex coupling parameters
that may be varied by slowly switching the laser field on and off. To
realize the generalization of the $U_2$ gate of Ref. \cite{duan01},
restrict to paths that correspond to the parametrization $\omega_0 =
\omega \sin \theta \cos \phi$, $\omega_1 = \omega \sin \theta \sin 
\phi$, and $\omega_a = \omega \cos \theta$. Under this condition, 
there is a pair of zero energy dark states of the form
\begin{eqnarray}
\ket{D_0(\theta,\phi)} & = & \cos \theta \cos \phi \ket{0} +
\cos \theta \sin \phi \ket{1} - \sin \theta \ket{a} ,
\nonumber \\
\ket{D_1(\theta,\phi)} & = & 
- \sin \phi \ket{0} + \cos \phi \ket{1} . 
\end{eqnarray}
For simplicity, in the adiabatic limit $\omega T \gg 1$, we may 
assume that the system's density operator diagonalizes in this 
dark state basis, i.e.,  
\begin{eqnarray}
\rho (\theta,\phi) & = & 
\frac{1}{2}(1+r) \ket{D_0(\theta,\phi)} \bra{D_0(\theta,\phi)} 
\nonumber \\ 
 & & + \frac{1}{2}(1-r) \ket{D_1(\theta,\phi)} \bra{D_1(\theta,\phi)}  
\end{eqnarray}
with $-1 \leq r \leq 1$. By using Eq. (\ref{eq:msconnection}), 
we obtain 
\begin{equation}
A = ir\sigma_y \cos \theta d\phi ,
\end{equation}
where $\sigma_x,\sigma_y,\sigma_z$ are the standard Pauli operators in
the $\ket{0},\ket{1}$ basis. However, this connection one-form is only
valid for a patch excluding the north and south pole of the sphere,
where the coordinates are not uniquely defined. This can be seen by
considering an infinitesimal loop $\delta {\cal C}$ around the north
pole, say, yielding
\begin{equation}
U[\delta {\cal C}] =
{\bf P} \exp \left( -\oint_{\delta {\cal C}} A \right) =
e^{-ir\sigma_y 2\pi}.
\end{equation}
To get around this problem we make the gauge transformation
\begin{equation}
A \longrightarrow A_N = V_NAV_N^{\dagger} + dV_NV_N^{\dagger}
\end{equation}
with
\begin{equation}
V_N = e^{-ir\sigma_y\phi} . 
\end{equation}
Explicitly,
\begin{equation}
A_N = -ir\sigma_y (1-\cos \theta) d\phi ,
\end{equation}
which is a valid gauge potential for the northern hemisphere.

Now, since $[A_{N;\phi} (\theta),A_{N;\phi} (\theta')]=0$ for 
any pair $\theta,\theta'$ along ${\cal C}$, path ordering is 
not needed and the mixed state holonomy becomes
\begin{equation}
U[{\cal C}] = e^{ir\Omega\sigma_y},
\end{equation}
where $\Omega$ is the solid angle enclosed by the path ${\cal C}$
in parameter space. In the pure state limit $r=1$, $U[{\cal C}]$
reduces to the $U_2$ gate in Ref. \cite{duan01}.

As mentioned above, we expect the mixing of the state to change around
the loop ${\cal C}$. To check this, let $\rho (0) =\frac{1}{2}(1+r)
\ket{0} \bra{0} + \frac{1}{2} (1-r) \ket{1} \bra{1}$, with purification
$\ket{\psi (0)} = \sqrt{\frac{1}{2}(1+r)} \ket{0} \otimes 
\ket{\varphi_0} + \sqrt{\frac{1}{2}(1-r)} \ket{1} \otimes 
\ket{\varphi_1} \equiv \ket{\psi_0}$. After having traversed 
the loop, we obtain
\begin{eqnarray}
\ket{\psi (T)} & = & U_{00} [{\cal C}] \ket{\psi_0} +
U_{01} [{\cal C}] \ket{\psi_1} 
\nonumber \\ 
 & = & \cos (r\Omega) \ket{\psi_0} + \sin (r\Omega) \ket{\psi_1}
\end{eqnarray}
with $\ket{\psi_1} = \sqrt{\frac{1}{2}(1+r)} \ket{1} \otimes 
\ket{\varphi_0} + \sqrt{\frac{1}{2}(1-r)} \ket{0} \otimes 
\ket{\varphi_1}$, which is a purification of 
\begin{eqnarray}
\rho (T) & = & \frac{1}{2} \Big( I + \sin \left( 2r\Omega \right)
\sigma_x + r\cos \left( 2r\Omega \right)
\sigma_z \Big) , 
\end{eqnarray}
whose Bloch vector traces out an ellipse in the $(x,z)$ plane with 
half axes $(1,|r|)$ when $\Omega$ varies. We obtain
\begin{eqnarray}
\Tr \rho^2 (T) & = &
\frac{1}{2} \Big( 1+r^2+(1-r^2) \sin^2 (2r\Omega) \Big) ,
\end{eqnarray}
which equals $\Tr \rho^2 (0) = \frac{1}{2}(1+r^2)$ if and only if 
$|r|=1$ (pure states) or $\Omega = n\pi/(2r)$, $n$ integer.

For $\Omega = \big( n+\frac{1}{2} \big) \pi/(2r)$, we obtain that
the initial mixed state $\rho(0)$ polarized along the $z$ axis has
transformed into the pure state $\rho(T) = \frac{1}{2} (I \pm
\sigma_x)$. This may be understood by looking at the final
purification, which for $n=0$ reads
\begin{eqnarray}
\ket{\tilde{\psi}(T)} & = & \frac{1}{\sqrt{2}} \Big( \ket{\psi_0} + 
\ket{\psi_1} \Big) = \frac{1}{\sqrt{2}} \big( \ket{0} + \ket{1}
\big) 
\nonumber \\ 
 & & \otimes \Big( \sqrt{\frac{1}{2} (1+r)} \ket{\varphi_1} +
\sqrt{\frac{1}{2} (1-r)} \ket{\varphi_2} \Big)
\end{eqnarray}
corresponding to a product state in ${\cal H}_S \otimes {\cal H}_A$. 
Conversely, one may transform a pure input state into a
mixed one using the mixed state holonomy $U[{\cal C}]$. To
illustrate this latter point, let $\ket{\psi'_0} =
\frac{1}{2} \sqrt{(1+r)} \big( \ket{0} + \ket{1} \big) \otimes 
\ket{\varphi_0} + \frac{1}{2} \sqrt{(1-r)} \big( \ket{0} - \ket{1}
\big) \otimes \ket{\varphi_1}$ and $\ket{\psi'_1} = \frac{1}{2}
\sqrt{(1+r)} \big( \ket{0} - \ket{1} \big) \otimes \ket{\varphi_0} +
\frac{1}{2} \sqrt{(1-r)} \big( \ket{0} + \ket{1} \big) \otimes 
\ket{\varphi_1}$, so that $\rho' (0) = \ket{0} \bra{0}$ is
obtained from the purification $\ket{\tilde{\psi}(0)} =
\frac{1}{\sqrt{2}} \big( \ket{\psi'_0} + \ket{\psi'_1} \big)$.
After having traversed the loop ${\cal C}$, we obtain
\begin{eqnarray}
\ket{\tilde{\psi} (T)} & = & \frac{1}{\sqrt{2}} \Big[ \big( U_{00}
[{\cal C}] + U_{10} [{\cal C}] \big) \ket{\psi'_0} 
\nonumber \\ 
 & & + 
\big( U_{01} [{\cal C}] + U_{11} [{\cal C}] \big)
\ket{\psi'_1} \Big]
\nonumber \\
 & = & \cos (r\Omega + \pi /4) \ket{\psi'_0} 
\nonumber \\ 
 & & + \cos (r\Omega - \pi /4) \ket{\psi'_1} ,   
\end{eqnarray}
which is a purification of the final density operator
\begin{eqnarray}
\rho' (T) = \frac{1}{2} \Big( I - r\sin (2r\Omega) \sigma_x + 
\cos (2r\Omega) \sigma_z \Big)
\end{eqnarray}
that is nonpure for $|r|<1$ unless $\Omega = n\pi /(2r)$, $n$ 
integer. Note that the accessible final states again constitute 
an ellipse in the $(x,z)$ plane, now with half axes $(|r|,1)$. 

More general manifolds of final states are obtained if the 
initial $S+A$ state is mixed. As an illustration of this, 
consider the input state 
\begin{eqnarray} 
\varrho (0) = \frac{1+R}{2} \ket{\psi_0} \bra{\psi_0} + 
\frac{1-R}{2} \ket{\psi_1} \bra{\psi_1}   
\end{eqnarray} 
with $-1 \leq R \leq 1$ and $\ket{\psi_0},\ket{\psi_1}$ defined 
above. This state evolves into $\varrho (T) = U[{\cal C}] 
\varrho (0) U^{\dagger} [{\cal C}]$, yielding the 
final density matrix of $S$ as 
\begin{eqnarray} 
\rho (T) = \frac{1}{2} \Big( I + R\sin \left( 2r\Omega \right)
\sigma_x + rR\cos \left( 2r\Omega \right)
\sigma_z \Big) , 
\end{eqnarray} 
which corresponds to a shrinking of the ellipse in the $(x,z)$ 
plane by a factor $|R|$. 

\section{Conclusions}
Holonomy effects for mixed states, as first conceived by Uhlmann
\cite{uhlmann86} and later reconsidered in Ref. \cite{sjoqvist00},
have recently attracted interest due to its potential importance
to fault tolerant quantum information processing. These previous
analyzes have mainly been concerned with the Abelian case, while,
on the other hand, universal sets of quantum gates must involve
noncommuting one- and two-qubit gates. In other words, to examine
the effect of nonpure input states in all-geometric quantum
information processing requires a concept of non-Abelian mixed
state holonomy.

In this paper, we have introduced such a concept of non-Abelian
holonomy for adiabatic transport of energetically degenerate mixed
quantal states. The holonomy effect may be understood as arising
in certain types of slowly changing Hamiltonians that couple the
considered system to an ancilla. 

The present type of holonomy effect has an inherent additional richness
compared to the standard Wilczek-Zee pure state holonomy in that it may
change the mixedness of the input state. We have explicitly demonstrated
that this feature may transform a mixed state into a pure one and
vice versa by purely geometric means. This provides a method for 
robust coherent control of quantal states, which may be of use 
for quantum information processing and communication purposes.

\section*{Acknowledgments}
The work by E.S. was supported in part by the Swedish Research Council.

\end{document}